\documentclass[12pt,letterpaper]{article}
\usepackage[a4paper, total={7in, 10in}]{geometry}

\usepackage{graphicx}
\usepackage{helvet}
\usepackage{authblk}
\usepackage{hyperref}
\usepackage{booktabs}
\usepackage{float}
\usepackage{caption}
\usepackage{subcaption}
\usepackage{tabularx}
\usepackage{longtable}
\usepackage{array}
\usepackage{amsmath}
\usepackage{amssymb} 
\usepackage{orcidlink} 
\usepackage{setspace}
\usepackage{makecell}
\usepackage[authoryear]{natbib}
\usepackage{tcolorbox}
\usepackage{verbatim}

\newcolumntype{Y}{>{\centering\arraybackslash}X}  

\makeatletter
\renewcommand{\maketitle}{\bgroup\setlength{\parindent}{0pt}
\begin{flushleft}
  \textbf{\@title}
  
  \@author
\end{flushleft}\egroup}
\makeatother

\title{\Large From Birdwatch to Community Notes, from Twitter to X:\\ four years of community-based content moderation\\ \vspace{.4 cm}}
\date{}

\author[1,2 \orcidlink{0000-0003-1912-8605}]{Saeedeh Mohammadi}
\author[3,4,8]{Narges Chinichian}
\author[4,8]{Hannah Doyal}
\author[2]{Anna Bertani}
\author[5]{Kristina Skutilova}
\author[2]{Hao Cui}
\author[2]{Michele d’Errico}
\author[6]{Siobhan Grayson}
\author[1,2,7,9\orcidlink{0000-0002-1800-6094}* ]{Taha Yasseri}

\affil[1]{School of Mathematics and Statistics, University College Dublin, Dublin, Ireland}
\affil[2]{School of Social Sciences and Philosophy, Trinity College Dublin, Dublin, Ireland}
\affil[3]{Institute for Theoretical Physics, Technical University of Berlin, Berlin, Germany}
\affil[4]{SPICED Academy, Berlin, Germany}
\affil[5]{School of Computer Science, University College Dublin, Dublin, Ireland}
\affil[6]{School of Sociology, University College Dublin, Dublin, Ireland}
\affil[7]{Faculty of Arts and Humanities, Technological University Dublin, Dublin, Ireland}
\affil[8]{These authors contributed equally}
\affil[9]{Lead contact}

\affil[*]{Correspondence: taha.yasseri@tcd.ie}

\begin{document}

\maketitle
\doublespacing
\section*{SUMMARY}

Community Notes (formerly known as Birdwatch) is the first large-scale crowdsourced content moderation initiative launched by X (formerly Twitter) in January 2021. As the Community Notes model gains momentum across other social media platforms, there is a growing need to assess its underlying dynamics and effectiveness. This paper provides a descriptive investigation of Community Notes during its first four years, examining its linguistic diversity, sourcing practices, Contributor activity, rating behaviour, and interaction networks. In addition, we release a curated dataset and accompanying source code to support future research, along with a review of prior research on Community Notes. We parsed Notes and ratings data from the first four years of the program and conducted language detection across all Notes. For English-language Notes, we extracted embedded URLs and identified discussion topics in each Note. Additionally, we constructed monthly interaction networks among the Contributors. Together, the descriptive analysis, dataset, code, and literature review provide a foundation for advancing research on Community Notes and community-based content moderation more broadly.

\section*{KEYWORDS}


Content Moderation, Community Notes, Birdwatch, Network Analysis, Topic Modelling

\section*{Introduction}

The rapid production and spread of user-generated content on social media platforms in the absence of editorial oversight have increased users’ exposure to false or misleading information \citep{thai2016big}.  In response, social media platforms have taken various approaches to content moderation. Content Moderation refers to the process of monitoring, flagging, or removing content that violates community guidelines or is deemed harmful. The main moderation strategies have been expert evaluation and automated systems. However, each comes with limitations. Expert evaluation, while often accurate, is costly and impractical at scale, given the sheer volume of content that needs moderation \citep{hassan2015quest}. Automated methods, on the other hand, are constrained by the biases and quality of their training data, often reinforcing systemic biases in classification \citep{binns2017like}. In light of these challenges, community-based content moderation, leveraging the collective judgment of users, has emerged as a promising alternative.

On January 23, 2021, X (formerly Twitter) launched Community Notes (formerly Birdwatch), the first large-scale community-driven initiative to moderate misleading content. To participate, Contributors must have been active on X for at least six months, have a verified phone number, and maintain a clean record of compliance with the rules \citep{XCommunityNotesSigningUp}. 

\begin{figure}
\centering
    \centering
     \includegraphics[width=\textwidth]{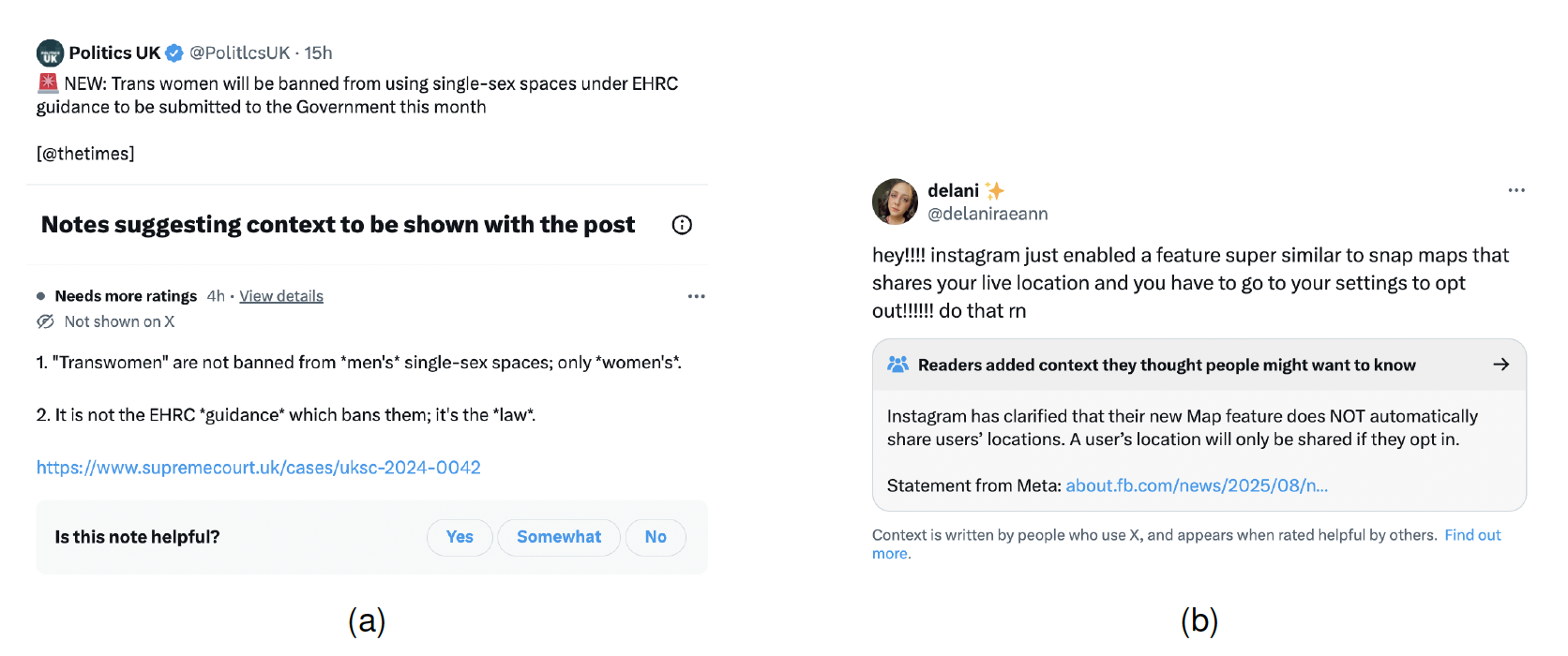}
    \caption{\textbf{Examples of Notes.} (a) A Note that requires more ratings and is shown only to Community Notes Contributors for additional rating. (b) A Note that has received the status of helpful and is publicly shown beneath a Post. }
    \label{fig:example}
\end{figure}

Contributors can write a Note to provide the missing context in relation to the Posts they find misleading. Other Contributors then rate these Notes as helpful, somewhat helpful, or not helpful (Figure~\ref{fig:example}(a)). 
The Note that receives the Helpful status based on the ratings from a diverse group of Contributors is displayed beneath the original Post in the timeline (an example is shown in Figure~\ref{fig:example}(b)) \citep{XCommunityNotesNotesOnTwitter}. 

Each Contributor has two impact metrics: writing impact and rating impact. The former increases when a Contributor’s Notes are consistently rated as helpful. The latter increases when the Contributor's rating of a Note aligns with the Note's eventual status determined by the rating algorithm, and decreases otherwise \citep{XCommunityNotesWritingAndRatingImpact}. New Contributors begin by rating existing Notes and can only write their own Notes once their rating score reaches a threshold of 5 \citep{XCommunityNotesWritingAbility}. Afterward, the Contributors can decide to write a Note on any Post. Furthermore, users on X can request a Note for a Post; if a Post receives enough requests, top writers, those with high writing impact, are notified to draft a Note for that Post \citep{XCommunityNotesNoteRequests}.

Community Notes has experimented with various rating systems and decision rules to determine which note to display for each post. The description above outlines only the core functionality of the system at the time of writing; we do not aim to detail specific algorithms and configurations, as these continue to evolve.

In the pilot version of Community Notes, Notes that received the highest number of helpful ratings were selected to appear in the timeline, regardless of the raters’ past behaviour. This was not immune to polarised rating among Contributors \citep{yasseri2023can} and partisan behaviour, such as Contributors rating Notes written by co-partisans as helpful and labelling Posts from cross-partisan Contributors as misleading \citep{allen2022birds}. To mitigate this issue, Community Notes adopted a {\it bridging algorithm} \citep{wojcik2022birdwatch} in the rating system. This approach places Contributors along an opinion spectrum based on their historical rating behaviour. A Note is deemed helpful only if it receives enough positive ratings from Contributors with opposing viewpoints. The goal of this algorithm is to elevate Notes rated helpful by a diverse set of Contributors. Community Notes is entirely open-source, with both the code and data publicly available \citep{communityNotesSource}.

In this work, we conduct a descriptive, observational investigation of Community Notes to characterise its functionality at scale. Through this analysis, we examine key aspects of how the system operates in practice and address the following questions:

\begin{itemize}
    \item \textbf{RQ1} Is Contributor attention evenly distributed across Posts and Notes, or is it concentrated on a small subset of content?
    
    \item \textbf{RQ2} What proportion of Notes reach a final status, and how long does it take for Notes to reach this status?
    
    \item \textbf{RQ3} What languages are used in Notes, and to what extent do Contributors engage across multiple languages?
    
    \item \textbf{RQ4} What types of external sources are cited in Notes, and which domains or categories of sources are most frequently referenced?

\item \textbf{RQ5} What are the main topics that the Notes are written about, and what temporal dynamics do the topics show?
    
\end{itemize}

To address these questions, we compile and analyse a comprehensive dataset of all Community Notes and their corresponding ratings over four years, from January 23, 2021, to January 23, 2025. We apply language detection to identify English-language Notes and, on this subset, perform topic modelling, extract URLs and domains, and construct interaction networks between Contributors based on their rating behaviour.

Beyond our empirical analysis, we position this work as a resource for the research community. We publicly release all processed data, including annotated Notes, extracted sources, and interaction networks, together with a literature review that provides an overview of existing work on Community Notes. While our analysis focuses on English-language content, the accompanying codebase is designed to be readily extended to multilingual settings.

\subsection*{Literature Review}

\paragraph{Comparative Performance of Community Notes.}

Community Notes represents a distinct approach to content moderation, differentiated not only by its crowdsourced structure but also by its epistemological foundation. As \citet{augenstein2025community} argue, Community Notes differs from traditional fact-checking in how it defines and validates corrective information. Whereas traditional fact-checking is often framed as the pursuit of an objective truth established by evidence, Community Notes relies on consensus among Contributors with different perspectives. This distinction has motivated a growing body of research examining the system’s functionality, credibility, and ethical implications.

A central question in this literature is how Community Notes performs relative to expert-led moderation and professional fact-checking. Existing findings suggest that Community Notes can be effective at identifying misinformation and flagging misleading content before expert fact-checkers in some cases. There is also substantial agreement between Community Notes Contributors and professional fact-checkers, with Contributor classifications often aligning closely with expert evaluations \citep{saeed2022crowdsourced}. However, the two systems appear to focus on different types of content. Contributors tend to prioritise Posts from influential users with larger followings, whereas experts are more likely to target content from less influential accounts \citep{drolsbach2023diffusion}. Compared to expert-led moderation, Community Notes is also more scalable and cost-effective \citep{martel2024crowds}.
 
A second comparison concerns how Community Notes performs relative to other crowdsourced fact-checking approaches. For example, in “Snoping”, users respond directly to Posts with fact-checks. Compared with Snoping, Community Notes Contributors tend to prioritise high-visibility content. When both systems evaluate the same content, which occurs relatively rarely, their assessments typically align \citep{pilarski2024community}. 

Community Notes also differs from other interventions in how users perceive and respond to it. Users generally view Community Notes more favourably than misinformation flags. Unlike binary labels that simply mark content as false, Notes provide explanatory context, which may foster greater trust and acceptance. \citet{drolsbach2024community} show that this contextual approach is more persuasive, as users tend to prefer interpretive frameworks over authoritative declarations of truth. Compared to traditional fact-checking interventions, Community Notes is less intrusive because users retain the autonomy to engage with or disregard the explanation \citep{augenstein2025community}. The system’s algorithmic transparency has also been shown to enhance user trust \citep{augenstein2025community}.

Despite these advantages, the literature identifies several limitations. Fact-checking complex or high-stakes Posts often requires domain-specific expertise that many Contributors may lack, and knowledgeable Contributors may not encounter the Posts for which their expertise is most relevant \citep{augenstein2025community}. Community Notes also operates more slowly than alternative crowdsourced approaches such as Snoping \citep{pilarski2024community}. A recent study analysing over 2.2 million Posts found that 99.3\% of misleading Posts received debunking replies within two hours of publication, whereas Community Notes took an average of 24.29 hours to appear beneath a Post \citep{zhang2025commenotes}. In addition, the system’s open design may leave it vulnerable to manipulation and adversarial attacks \citep{saeed2022crowdsourced}. Finally, Community Notes remains connected to the broader professional fact-checking ecosystem: one in twenty Notes explicitly references fact-checking sources, a proportion that increases for sensitive topics \citep{borenstein2025can}. These findings suggest that while Community Notes offers a scalable and less intrusive alternative to traditional moderation, the production of high-quality Notes may still depend in part on external expertise and institutional fact-checking infrastructure.

\paragraph{Impact on Engagement and User Behaviour.} While the studies above demonstrate that Community Notes can accurately identify misleading content and produce high-quality Notes, a critical question remains: can Community Notes reduce user engagement with such content? Results from A/B testing indicate that users exposed to Community Notes are 25–34\% less likely to like or share flagged Posts \citep{wojcik2022birdwatch}. Additional studies report that Posts with attached Notes have higher deletion rates and that attached Notes reduce reposts. This effect is more pronounced for Posts containing embedded media than for text-only Posts \citep{chuai2026community}. Similarly, \citet{drolsbach2023diffusion} estimate that misleading Posts with attached Notes receive 36.85\% fewer reposts than non-misleading ones. Beyond reposts, annotated Posts tend to accumulate fewer likes, views, and replies overall. Moreover, when Notes appear beneath a Post, it spreads less widely and deeply across the platform \citep{slaughter2025community}. 

Engagement with misleading Posts also depends on how users perceive their believability and harmfulness. \citet{drolsbach2023believability} analysed contributor responses to questions assessing whether misleading posts were perceived as believable and/or harmful. By linking these perceptions to the repost count, they found that misleading posts considered easily believable but not particularly harmful were more likely to be reposted. 

Community Notes also shape how readers respond to misinformation.  Studies show that once a Note is displayed underneath a Post, there is a measurable shift in users’ emotional responses, including a 7.3\% increase in negativity in replies. More specifically, displayed Notes cause a 13.2\% increase in anger, a 4.7\% increase in disgust, and a 16.0\% rise in moral outrage in replies. These effects are more pronounced for political content than for non-political content \citep{chuai2025community}. \citet{kankham2024community} find that Notes are particularly effective in reducing belief and the spread of “wish” rumours (misinformation aligned with users’ desires), whereas presenting related news articles is more effective in countering “dread” rumours (misinformation that evokes fear).

Community Notes influence not only how users engage with content but also how they post on the platform. Studies show that users whose content receives Notes tend to post less frequently overall; however, their subsequent Posts display greater cognitive processing. Exposure to Community Notes also increases the reliability of users’ own writing while reducing extreme sentiment. These cognitive benefits, however, come with a trade-off as users exposed to Notes tend to participate less actively on the platform \citep{borwankar2022democratization}. Being fact-checked through Community Notes does not appear to meaningfully reduce an author’s follower count, suggesting that the system’s influence on social connections remains limited \citep{bobek2026community}. Receiving a Note can also have the unintended effect of increasing a user’s visibility: users, particularly those with smaller followings, who are “noted” often gain additional followers, potentially amplifying their reach rather than diminishing it \citep{wirtschafter2023future}. 

\paragraph{Temporal Dynamics and System Efficiency.} Although Community Notes can reduce engagement once displayed, recent studies indicate that its overall effect is limited by delays in Notes reaching “helpful” status \citep{chuai2024did, bak2023limiting}. \citet{renault2024collaboratively} report that while annotated Posts experience nearly a 50\% drop in reposts and over a 30\% reduction in replies and quotes, this effect is highly temporal. On average, it takes 15 hours for a Note to be published, by which time a Post has typically reached 80\% of its total audience. Similarly, \citet{chuai2026community} show that by the time a Post reaches its half-life, only 13.5\% of helpful Notes have been attached. \citet{de2025supernotes} further find that in 91\% of Posts where at least one Note was proposed, none ultimately reached “helpful” status. Relatedly, most Contributors have yet to produce a single Note rated as helpful \citep{wirtschafter2023future}. These publication barriers may also affect Contributor retention. \citep{arjmandi2025threats} show that having a Note published increases the retention of first-time Contributors by 5\%, implying that low publication rates may threaten the long-term sustainability of the Community Notes ecosystem.

These delays are particularly problematic in time-sensitive or politically charged contexts. For instance, one report found that 74\% of accurate Notes related to the 2024 U.S. presidential election were never shown to users, allowing misleading Posts without a Community Note to spread 13 times faster than those with a Community Note \citep{CCDH2024}. Researchers attribute this to the design of the rating algorithm: the more polarising the content (e.g., national elections), the less likely accurate Notes are to receive “helpful” status \citep{bouchaud2025algorithmic}.

\paragraph{Contributor Behavior, Bias, and Platform Effects.} 
Contributors select which Posts to write Community Notes for based on multiple factors. A recent study shows that among 90,000 Posts for which X users requested a Note, the Posts perceived as more misleading by GPT-4.1 and the Posts by authors who are more frequently fact-checked were more likely to attract Notes \citep{chuai2026request}. Political partisanship plays a central role in how Contributors both select and evaluate Notes. Research shows that partisanship is often a stronger predictor of judgment than the content itself. Contributors are more likely to evaluate Posts from political opponents negatively and to rate Notes written by Contributors of opposing affiliations as unhelpful, thereby reinforcing existing ideological divides \citep{allen2022birds}. Network analyses of positive interactions reveal that the initial rating system, where the Note with the highest number of helpful ratings was displayed, did not promote cross-partisan interaction. Instead, Contributors clustered into polarised groups \citep{yasseri2023can}. This pattern mirrors broader trends of political polarisation observed on social media. Studies of signed interaction networks in Community Notes reinforce this finding, showing that Contributors consistently form clusters based on shared political ideologies \citep{fraxanet2024unpacking}. Topic modelling and network analyses further demonstrate that such polarisation is especially pronounced in political discussions, whereas evaluations of non-political Posts display substantially less ideological bias \citep{champaigne2022birdwatch}. 

Recent work suggests that the delays in Note publication also stem from such behaviour. \citet{truong2025community} exposed the algorithm to more than 15,000 simulated datasets with varying attributes and found it highly sensitive to rater biases, such as a tendency to downrate Notes written by Contributors of the opposite political persuasion. Their results show that even a small fraction of low-quality raters can suppress helpful Notes to the extent that they never reach publication. 

Despite these limitations, evidence suggests that diversity among fact-checkers can enhance fact-checking outcomes. Studies show that political motivation can help address a key challenge in volunteer-based moderation systems—insufficient content flagging. Contributors with strong political motives flag twice as many Posts as less politically motivated individuals, without any notable decline in the quality of their flags \citep{martel2025political}. Moreover, research comparing individual tagging (e.g., labelling misinformation independently) with Community Notes finds that the latter exposes users to more diverse perspectives. Unlike individual tagging, which can deepen echo chambers, Community Notes’ peer-review structure temporarily increases informational diversity \citep{kim2025differential}. 

Platform-level changes in Community Notes have also shaped Contributor behaviour over time. After the introduction of a privacy-preserving policy that anonymised Contributor identities—preventing their Community Notes activity from being linked to their X profiles—Contributors began posting more frequently, and the neutrality and overall quality of their Posts improved \citep{borwankar2024unveiling}. The system’s expansion has also influenced Contributors' behaviour over time. Early in the program, most Notes were written on Posts from users estimated to lean liberal \citep{wirtschafter2023future}. However, as the Contributor base broadened, the partisan distribution shifted. \citet{renault2025republicans} found that Posts from Republican users were more likely to be flagged with Community Notes.

Some evidence suggests that Contributors also use Community Notes as a space for political deliberation. The system exhibits distinctive patterns when Contributors write Notes asserting that a Post should not be considered misleading. These Notes often begin with “NNN”, meaning “Note Not Needed”. \citet{razuvayevskaya2025timeliness} argue that such cases reflect a repurposing of the platform for debate rather than moderation. Their study shows that Posts receiving an “NNN” Note are more likely to yield a published Note than Posts that do not, suggesting that Contributors use the system not only to correct misleading content but also to contest whether moderation is warranted. This deliberative function connects Community Notes to broader work on collective moderation systems. Research on deliberative platforms such as Polis similarly suggests that structured participation can improve the quality of public discourse and support misinformation detection by promoting reflection and consensus \citep{megill2022coherent}.

However, this participatory structure also creates risks for Contributors. These include system manipulation, exploitation of unpaid labour, and the marginalisation of underrepresented voices \citep{augenstein2025community}. Furthermore, some features—such as prompting Contributors to revisit flagged content—may unintentionally reinforce exposure to misinformation \citep{wang2024efficiency}. Other ethical concerns have also been raised about the open structure of Community Notes, particularly the potential psychological harm to Contributors who encounter harmful or sensitive content \citep{augenstein2025community}. 

\paragraph{Note Characteristics and Sources.} Several studies have examined the characteristics of Notes produced in Community Notes to better understand what makes a Note more effective or persuasive. Research indicates that Notes on Posts that are flagged as misleading tend to be longer, more complex, and more negatively worded; traits that may make them more informative but potentially less accessible to general audiences \citep{prollochs2022community}. The emotional tone of a Note also plays a key role: while emotionally charged language can increase engagement, it tends to reduce perceived trustworthiness. On average, Notes with high emotional content are rated as less helpful than those adopting a more neutral tone \citep{phillips2025emotional}. Moreover, Notes that closely align with the topic of the associated Post receive higher helpfulness ratings \citep{simpson2022obama}.

Citing sources is another critical element of Note quality. Contributors are encouraged to support their claims with external references, and including a source increases the likelihood of a Note reaching “helpful” status by a factor of 2.33. However, sources perceived as politically biased reduce the likelihood that a Note will be rated as helpful \citep{solovev2025references}. \citet{kangur2026checks} analysed citation patterns and found that the most frequently referenced sources were X and Wikipedia, with a noticeable left-leaning bias. Notes that cited more balanced or factually rigorous sources received higher helpfulness scores.

\paragraph{Proposed System Improvements.} Several studies have proposed modifications to Community Notes to enhance its effectiveness. \citet{lloyd2026beyond} argue that future research should focus on how the system’s design and implementation influence its efficiency, including aspects such as Contributor interaction, Note presentation, and the integration of technological advances like Large Language Models (LLMs). Building on this, various algorithmic and hybrid approaches have been explored.

One such approach is HawkEye, a graph-based algorithm that iteratively evaluates the quality of Contributors, Notes, and Posts. By applying scoring principles and smoothing techniques to handle sparse data, HawkEye improves the consistency and robustness of evaluations \citep{mujumdar2021hawkeye}.

Supernotes adopt a different approach by leveraging LLMs to generate Notes that are more likely to be rated as “helpful” than those written by individual Contributors. In this framework, an LLM reads all existing Notes, generates multiple candidate Notes, and evaluates them with a Personalised Helpful Model (PHM), which estimates the probability of receiving a “helpful” rating from a synthetic jury of diverse raters. The highest-scoring Note is then presented as the Supernote, and empirical evidence shows that Contributors rate these as significantly more helpful than the best human-written Notes \citep{de2025supernotes}.

Additionally, \citet{wu2025beyond} proposed CrowdNotes+, a similar hybrid framework tailored to health-related content. In this model, LLMs can either autonomously write Notes or use evidence gathered by human Contributors. Results show that LLMs produce more accurate and contextually balanced Notes than humans do, and, when sourcing their own evidence, they tend to select higher-quality material.

Other scholars extend this idea into a human–AI hybrid framework in which both human Contributors and LLMs can propose Notes, but only humans serve as raters and evaluators. This division of labour preserves the human judgment while enabling LLMs to accelerate the generation of high-quality Notes \citep{li2025scaling}. Pushing this idea further, \citet{costabile2025assessing} simulated crowds using generative agents with diverse demographics and ideological perspectives. These agents outperformed human crowds in truthfulness classification, demonstrated higher internal consistency, were less susceptible to social and cognitive biases, and relied more on systematically informative criteria.

Another line of research emphasises diversity and collaboration in the Note-writing process rather than focusing solely on the rating stage \citep{yasseri2023can}. In an online experiment, \citet{juncosa2026benefit} found that duos produced more helpful Notes than individuals, with diverse duos performing better on Republican-leaning Posts. Building on this insight, \citet{mohammadi2025ai} used LLMs to synthesise political diversity during Note writing. In a study involving over 800 participants, Contributors who received argumentative feedback from an LLM were more likely to produce higher-quality Notes. 

\section*{Results}
\subsection*{Overview}

\begin{figure}
    \centering\includegraphics[width=0.85\textwidth]{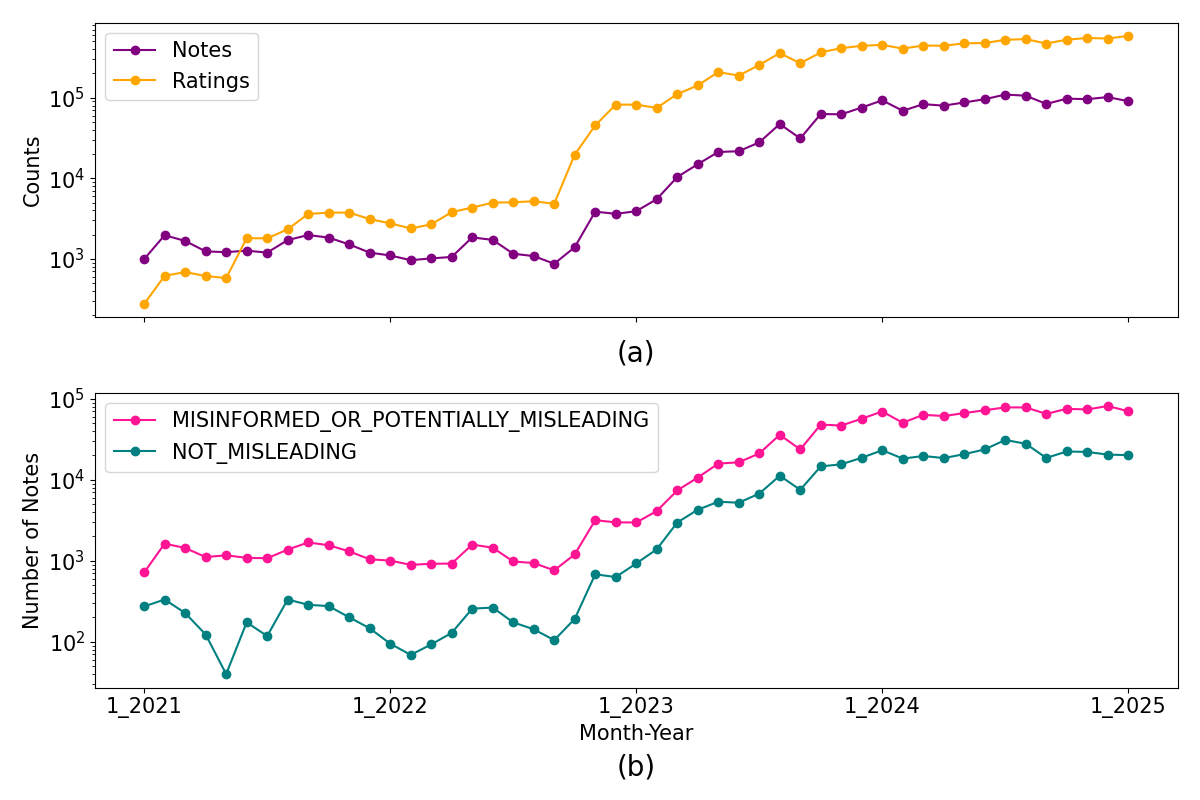}
    \caption{\textbf{Classification and the number of Notes per month.} 
    (a) The number of Notes and ratings created within each month on a log scale. (b) The monthly number of Notes that classify the Posts as ``potentially misleading or misinformed'' and ``not misleading'' over four years on a log scale.}
    \label{fig:note_class}
\end{figure}

Community Notes allows Contributors to classify a Post as either misleading or not misleading. The accompanying Note provides the rationale for this classification and provides additional context for the Post. Figure~\ref{fig:note_class} (b) shows the overall volume of Notes tagged as misleading or not misleading over the four-year period.

\begin{figure}
    \centering
    \includegraphics[width = 0.95\textwidth]{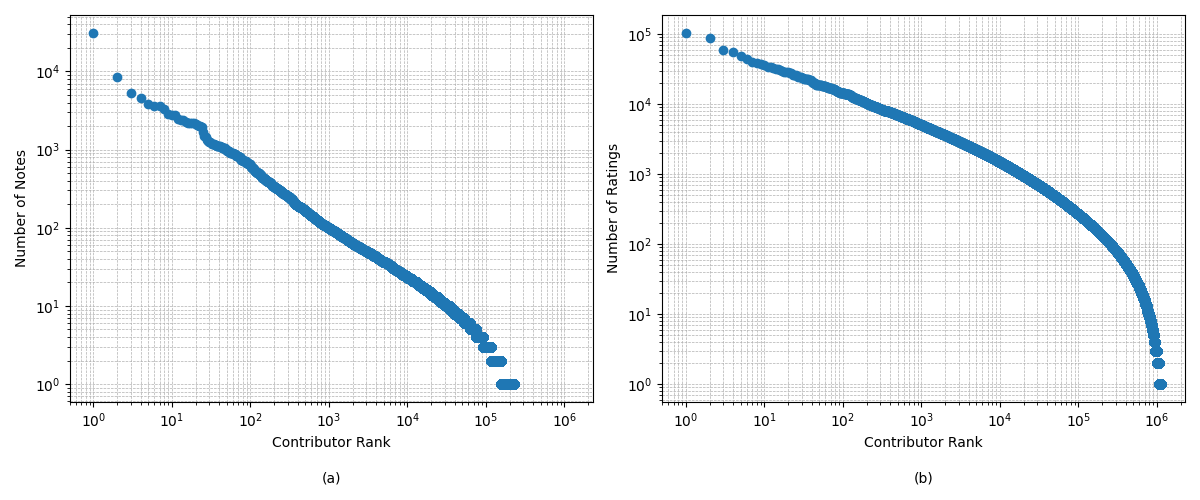}

    \caption{\textbf{The distribution of Notes and ratings of Contributors.}(a) Log-log rank-plot showing the number of Notes written by each Contributor vs their rank. (b) Log-log rank-plot of the number of ratings committed by each Contributor vs their rank.
    \label{fig:user-ranks}
    }
\end{figure}

Since its launch in January 2021, Community Notes has accumulated a substantial volume of contributions. 227,702 unique Contributors have written 1,614,743 Notes. However, this activity is not evenly distributed across Contributors. A small number of Contributors are responsible for a large share of the Notes. One Contributor alone has authored 33,186 Notes, an account that appears to be automated, consistently flagging impersonation attempts related to Non-Fungible Tokens (NFTs) and cryptocurrency scam accounts. The distribution of Notes authored per Contributor is highly skewed, as illustrated in Figure~\ref{fig:user-ranks}(a).

A similar pattern appears in the distribution of ratings. Figure~\ref{fig:user-ranks}(b) presents a rank plot of the number of ratings submitted by each contributor. The distribution again reveals substantial concentration in rating activity, with a small group of contributors accounting for a disproportionately large share of all ratings.

\begin{figure}
    \centering
    \includegraphics[width = 0.95\textwidth]{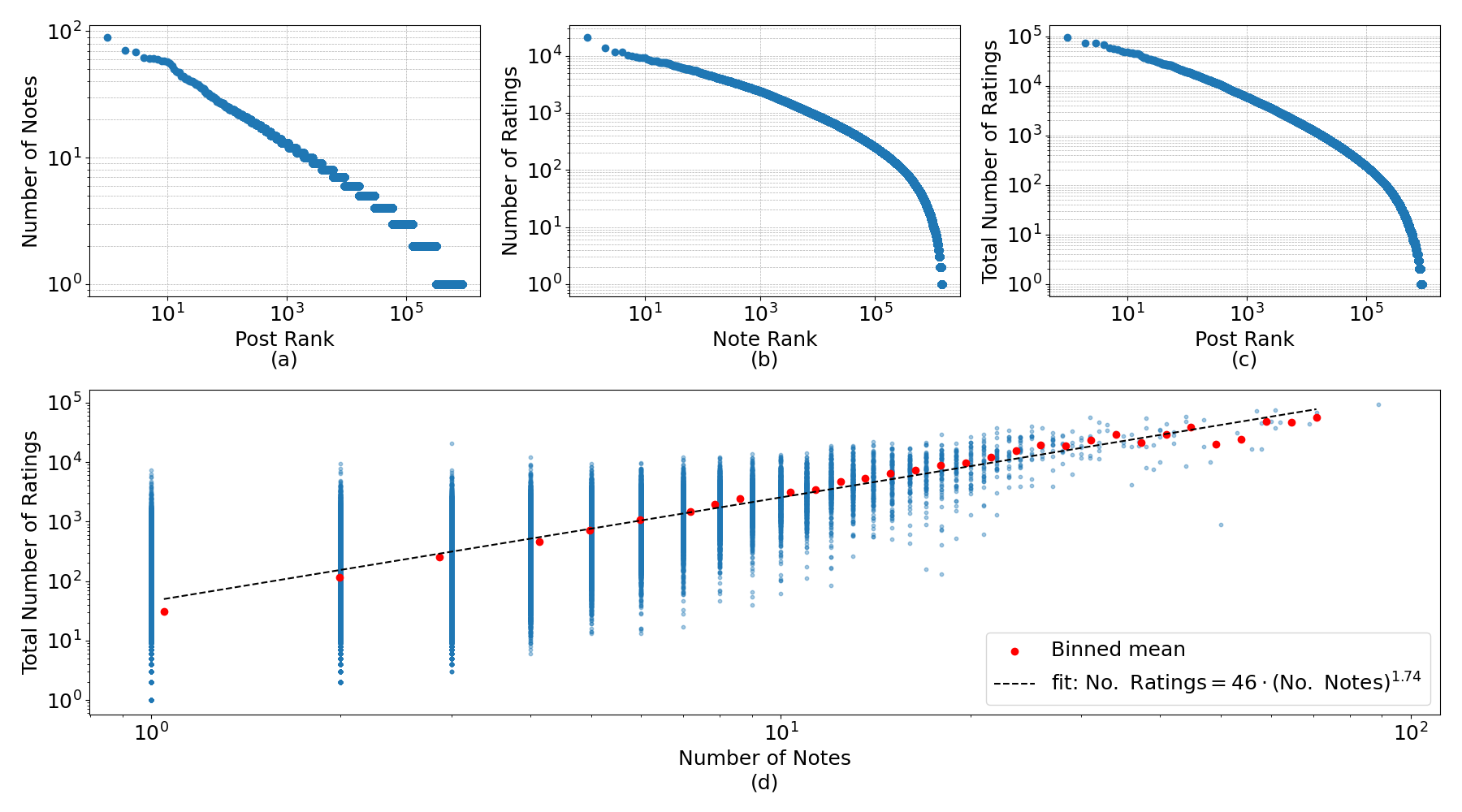}
\caption{\textbf{The distribution of Notes and ratings of Posts.} 
(a) Log-log rank plot of the number of Notes written on each Post. 
(b) Log-log rank plot of the number of ratings each Note has received. 
(c) Log-log rank plot of the total number of ratings on all Notes written on each Post. 
(d) Scatter plot (log-log) of the total number of ratings on all Notes written on each Post versus the number of Notes written on that Post. The red dots represent the binned mean number of ratings for a given range in the number of Notes. The black dashed line shows the fitted power-law relationship $y=A\cdot x^b$, where $A = 46$ is the scaling coefficient and $b = 1.74$ is the exponent.
    \label{fig:eda_new}
}
\end{figure}

The Notes in our dataset were written on 1,016,673 distinct Posts. As shown in Figure~\ref{fig:eda_new}(a), the distribution of Notes per Post is highly skewed: most Posts receive only a few Notes, while a small number attract substantial attention. The most annotated Post in the dataset received 90 Notes. This Post, made by Donald Trump on August 25, 2023, includes his mugshot alongside the caption: ``Election Interference, Never Surrender! DonaldjTrump.com''. As of this writing, despite the large number of Notes on this Post, none of them has reached ``helpful'' status, and therefore, no Note appears beneath it on the platform. 

Highly visible Posts can therefore attract multiple competing Notes without necessarily receiving a ``helpful'' Note. Although the number of ratings tends to increase with the number of notes, as shown in Figure~\ref{fig:eda_new}(d), many highly noted posts still do not have a displayed Note.

\subsection*{Content}
\begin{figure}
    \centering
    \includegraphics[width=0.8\textwidth]{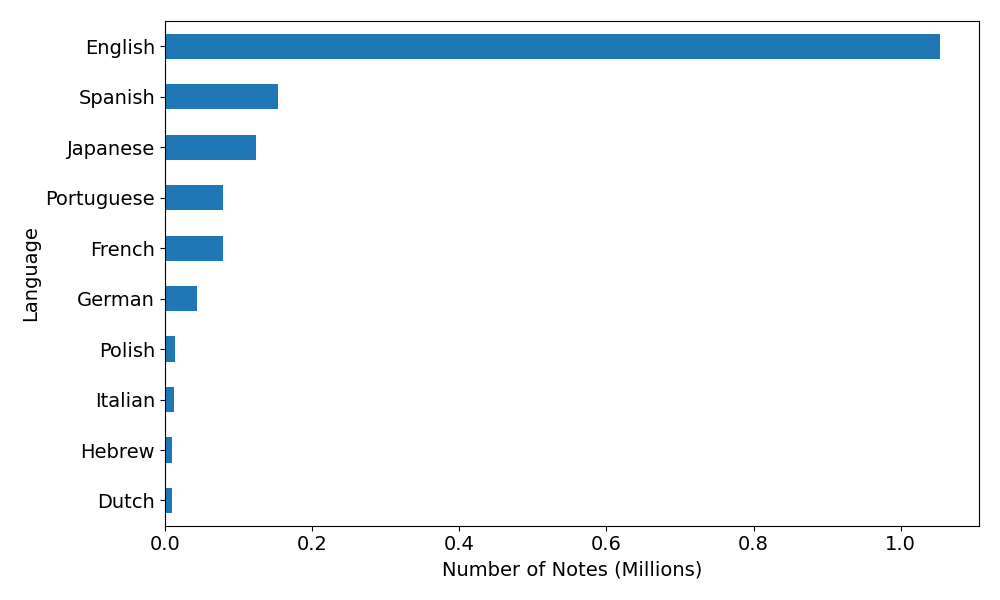}
    \caption{\textbf{Top 10 languages in Community Notes.} The number of Notes written in each of the top 10 languages used in Community Notes.}
    \label{fig:langs}
\end{figure}

Community Notes has been made available to X users in over 60 countries, with Notes written in 103 different languages. Figure~\ref{fig:langs} shows the top 10 languages in the dataset, along with the number of Notes written in each.
The vast majority of Contributors write in only one language: only 35,515 Contributors, approximately 16\% of all Note authors, have written Notes in more than one language. Even among the most multilingual Contributors, Notes are typically written in a single language. For example, the Contributor who used the most distinct languages has written Notes in 17 languages. Of their 500 Notes, 412 are in English, while most of their Notes in other languages consist of just a single Note each.

Given the prominence of English among Community Notes (Figure~\ref{fig:langs}), the remainder of our analysis focuses on English-language Notes. Among these Notes, 79.6\% contain at least one URL, reflecting the platform’s emphasis on source-backed claims. While drafting a Note, Contributors are instructed to include links to external sources and are prompted to indicate whether they believe the source would be considered trustworthy by most people.

We classified the 30 most frequently cited domains by political leaning using Ad Fontes Media’s 2023 ratings. Ad Fontes assigns each news source a score along a left–neutral–right spectrum (Left: –42 to –10, Neutral: –9 to +9, Right: +10 to +42). These ratings are determined by panels of trained analysts with diverse political ideologies who evaluate samples of articles from each source and manually assign bias scores based on language, framing, and political positioning \citep{adFonts}. 

\begin{figure}
    \centering
    \includegraphics[width=0.9\textwidth]{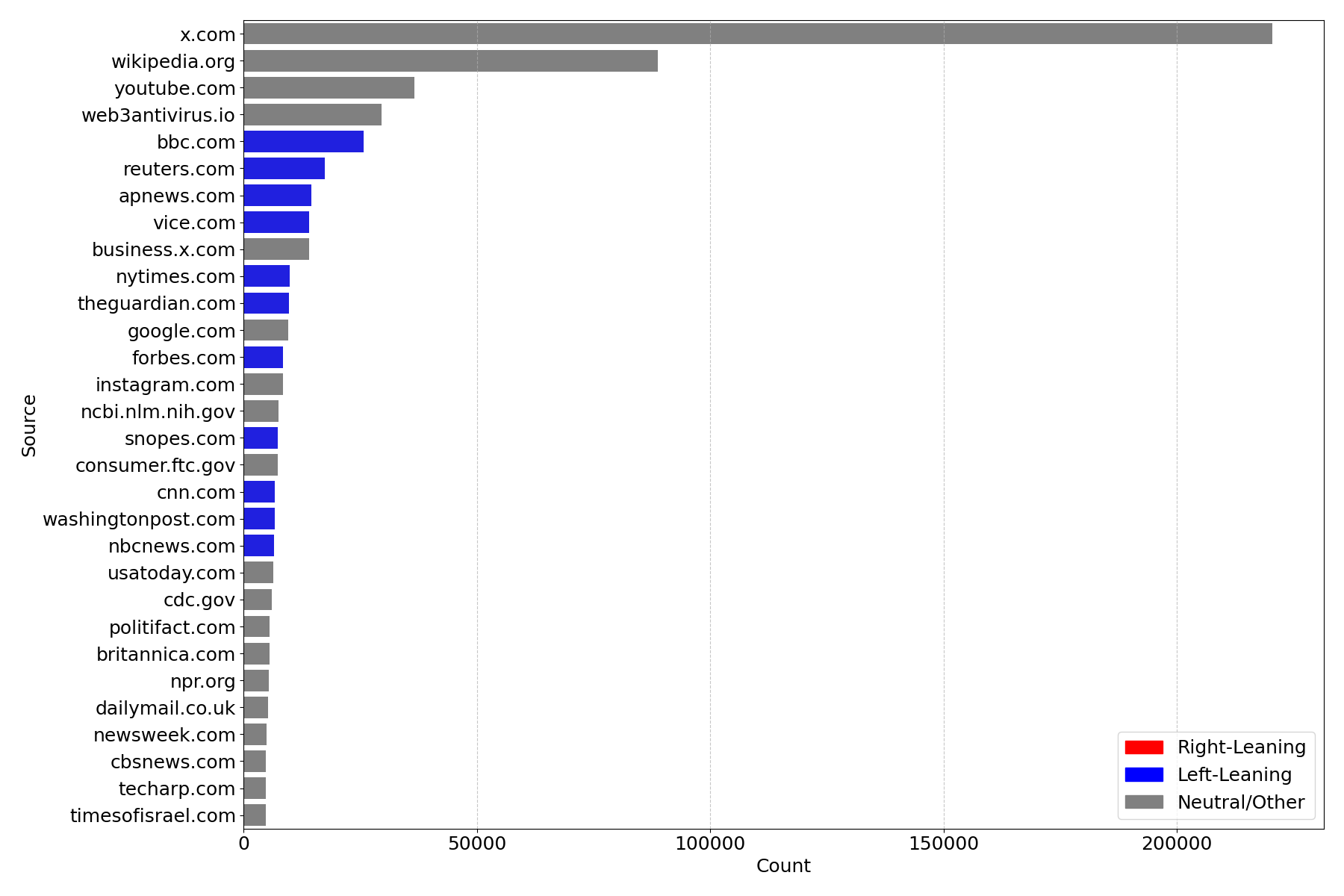}
    \caption{\textbf{Top 30 sources cited in Community Notes.} Top 30 cited domains in Community Notes, coloured by their (U.S.-based) political leaning. The X-axis shows the number of times each domain has been referenced in the Notes.}
    \label{fig:note_leaning}
\end{figure}

Figure~\ref{fig:note_leaning} displays the top 30 cited domains and the number of times each was referenced. Crowd-sourced platforms such as Wikipedia, X, and YouTube rank among the most frequently cited. The bar colours represent the (U.S.-based) political leaning of each domain, revealing that most cited sources are classified as neutral, although several left-leaning domains also appear.

\begin{figure}[htp!]
    \centering
    \includegraphics[width=\textwidth]{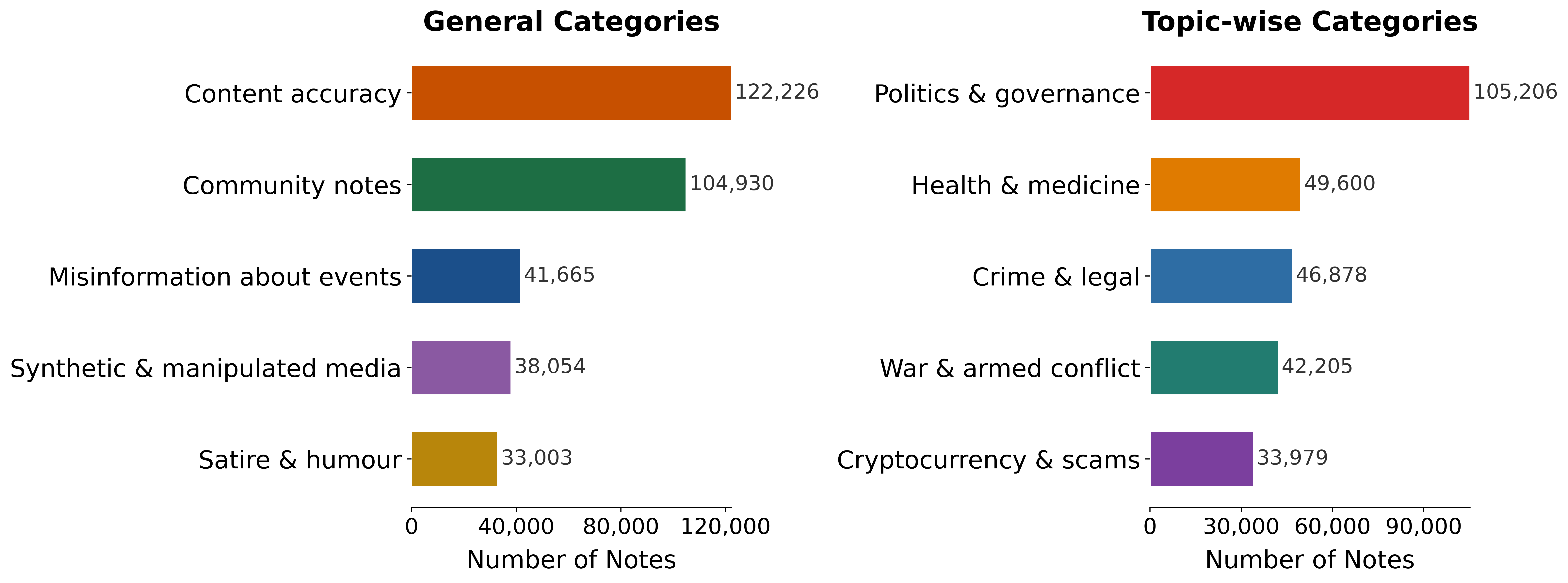}
    \caption{\textbf{Distribution of Community Notes across general and topic-wise categories.} 
    The left panel shows the five most frequent \textit{general} categories, which capture the 
    functional nature of the note (e.g., whether it corrects factual inaccuracies or flags 
    synthetic content). The right panel shows the five most frequent \textit{topic-wise} 
    categories, reflecting the domain of the annotated post. Categories were 
    derived from a two-stage taxonomy construction process, yielding a final taxonomy of 42 
    categories. Bars are sorted by frequency in descending order; values indicate the absolute 
    number of notes assigned to each category.}
    \label{fig:top10_categories_bar}
\end{figure}

\begin{figure}[htp!]
    \centering
    \includegraphics[width=\textwidth]{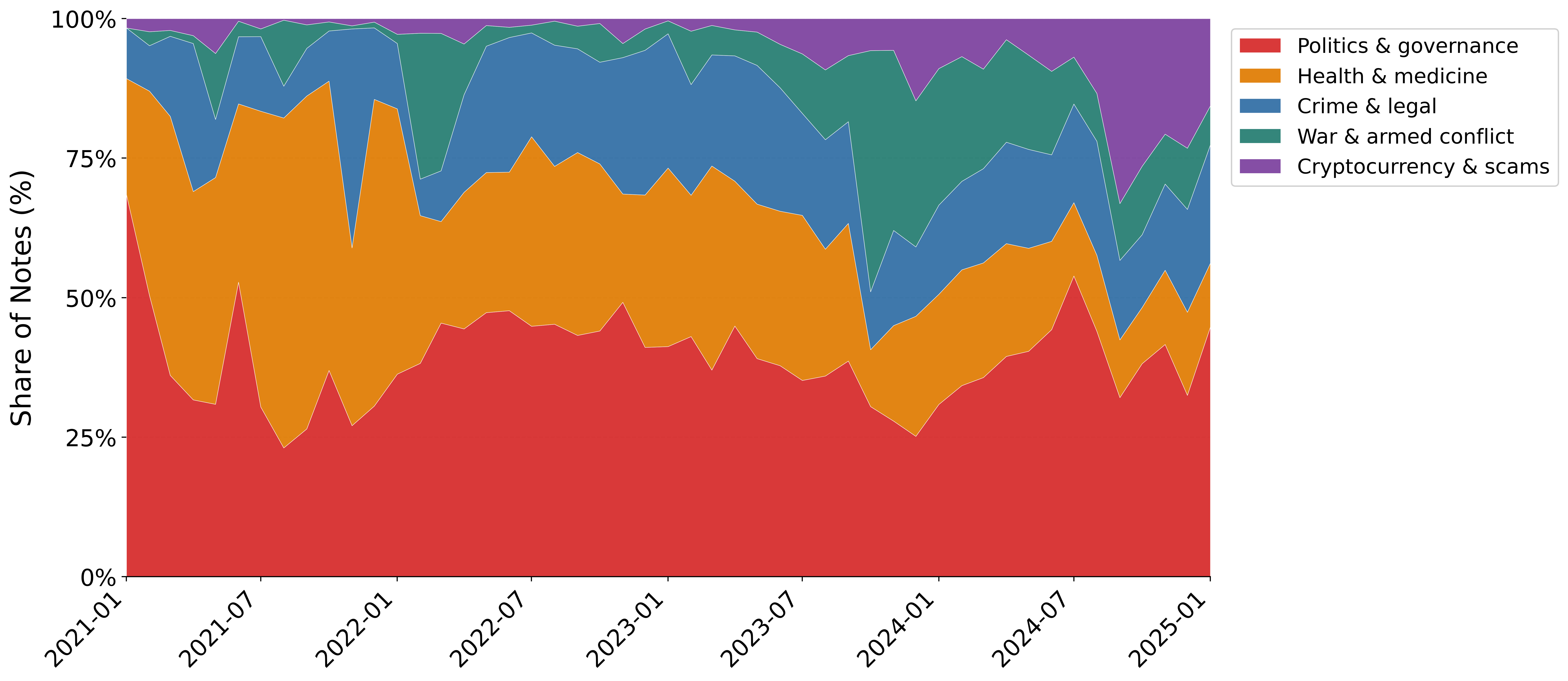}
    \caption{\textbf{Monthly share of Community Notes across the five most frequent 
    topic-wise categories (2021--2025).} Each band represents the proportional share 
    of notes assigned to a given category in a given month, normalised to 100\% to 
    highlight shifts in the relative composition of note-writing activity over time. }
    \label{fig:streamgraph_topic}
\end{figure}

Figure~\ref{fig:top10_categories_bar} shows the distribution of Community Notes 
across general and topic-wise categories. Among general categories, \textit{Content 
accuracy} is the most frequent (n\,=\,122,226), encompassing notes that directly 
correct factual errors in posts, for instance, clarifying misrepresented statistics, 
debunking false claims about public figures, or providing accurate scientific context. 
\textit{Community notes} (n\,=\,104,930) constitutes the second largest general 
category and includes meta-notes, often arguing that a note is unnecessary or biased (e.g., 
\textit{``NNN. Not Note Needed''}). 
\textit{Misinformation about events} (n\,=\,41,665) covers notes correcting false 
or misleading descriptions of specific real-world events, such as misattributed footage 
or incorrect timelines. \textit{Synthetic \& manipulated media} (n\,=\,38,054) flags 
AI-generated or digitally altered images and videos, while \textit{Satire \& humour} 
(n\,=\,33,003) labels notes clarifying that the content is satirical or comedic rather than factual.

Among topic-wise categories, \textit{Politics \& governance} is the most frequent 
(n\,=\,105,206), reflecting notes addressing political claims, electoral misinformation, 
and statements by public officials. \textit{Health \& medicine} (n\,=\,49,600) is the second 
most common, covering vaccine misinformation, false medical claims, and misleading 
health statistics, including notes addressing COVID-19 vaccine safety, cancer treatment 
claims, and drug overdose data. \textit{Crime \& legal} (n\,=\,46,878) includes notes 
clarifying legal proceedings, court verdicts, and law enforcement actions. \textit{War \& armed conflict} 
(n\,=\,42,205) encompasses notes related to ongoing conflicts, casualty figures, and 
the attribution of military actions, with a notable concentration of notes addressing 
the Israel-Gaza conflict and the war in Ukraine. Finally, \textit{Cryptocurrency \& 
scams} (n\,=\,33,979) captures notes warning about fraudulent investment schemes, 
phishing links, and fake token promotions.

Figure~\ref{fig:streamgraph_topic} illustrates how the proportional composition of 
note-writing activity shifted over the observation period (January 2021 -- January 2025). 
\textit{Politics \& governance} consistently represents the largest share of notes 
throughout the entire period, accounting for approximately 30--45\% of monthly activity. 
\textit{Health \& medicine} was mainly prominent in the early period 
(2021--2022), likely reflecting note-writing activity related to the COVID-19 
pandemic and associated vaccine discourse. Its relative share declined progressively 
from 2022 onward as other thematic domains gained prominence. \textit{War \& armed 
conflict} shows a marked increase from late 2023, coinciding with the escalation of 
the Israel-Palestine conflict and continued coverage of the war in Ukraine. 
\textit{Cryptocurrency \& scams} emerged as a visible category from 2023 onward, 
consistent with the broader growth of cryptocurrency-related misinformation during 
this period. \textit{Crime \& legal} maintained a relatively stable share throughout 
the observation window, with modest growth toward the end of the period.

\subsection*{Interactions}
\begin{figure}
\centering
\includegraphics[width=0.85\textwidth]{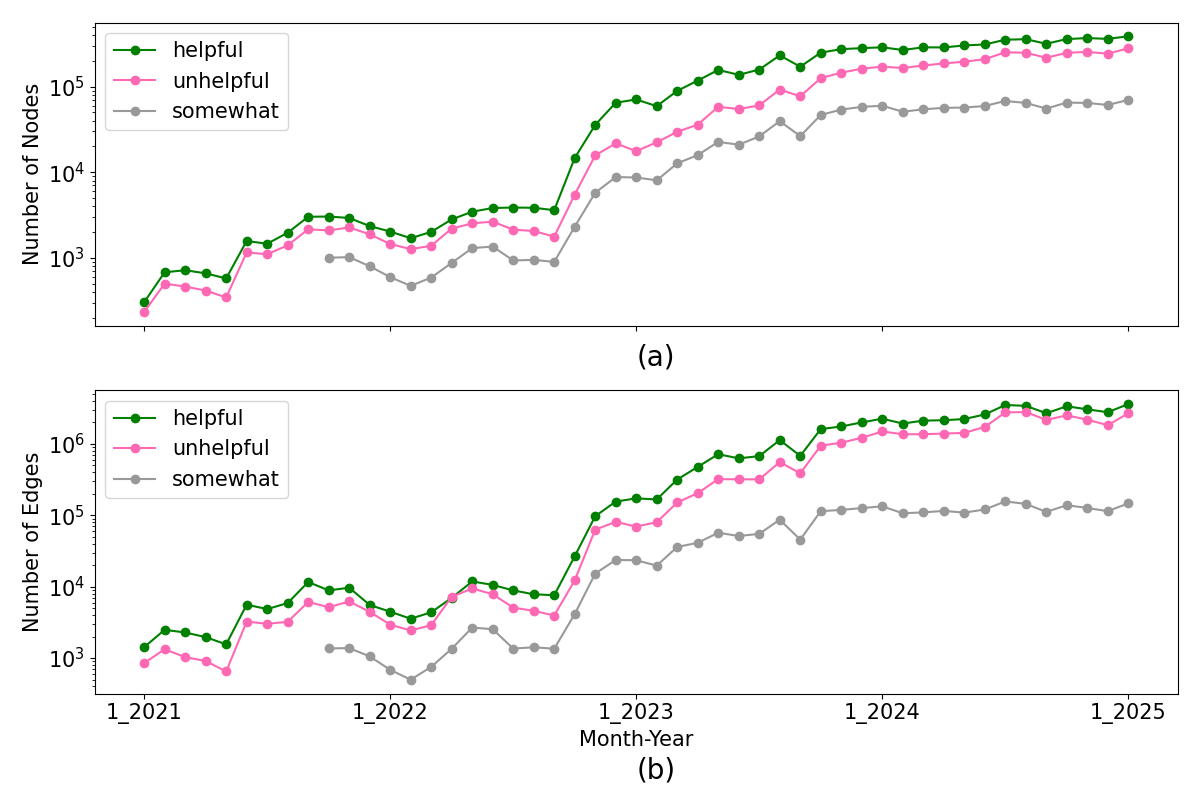}
\caption{\textbf{The size of the ratings network over time.} (a) The number of nodes for the helpful, unhelpful, and somewhat helpful networks for each month on the log scale. (b) The number of edges for the helpful, unhelpful, and somewhat helpful networks for each month on the log scale. Since the ``somewhat helpful'' option was introduced only at the end of June 2021, the somewhat helpful network did not exist before then.}
\label{fig:no_nodes_edges}
\end{figure}

We constructed monthly rating networks covering the full four-year dataset. In each network, nodes represent Contributors, and directed edges represent ratings: an edge from Contributor A to Contributor B indicates that A rated a Note written by B. The edge weight corresponds to the number of such ratings. For each month, we generated three separate networks: one for helpful ratings, one for somewhat helpful ratings, and one for unhelpful ratings. Figure~\ref{fig:no_nodes_edges} shows the number of nodes and edges in each of these networks over time.

\begin{figure}[htp!]
    \centering
    \includegraphics[width = \textwidth]{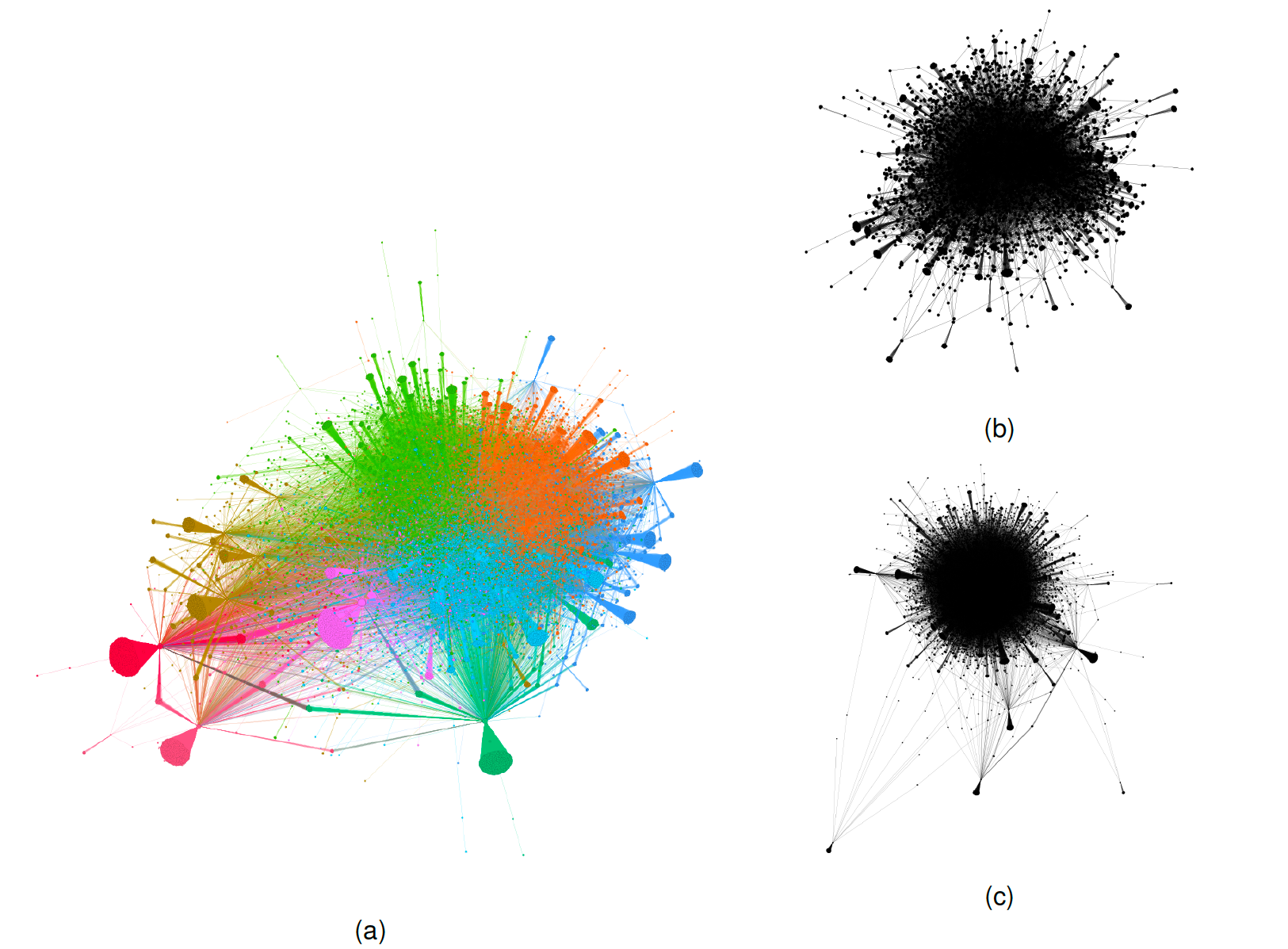}
   \caption{\textbf{Networks of January 2023.} The network visualisation of (a) helpful ratings, (b) somewhat helpful ratings, and (c) unhelpful ratings made in January 2023.
    Node colours indicate cluster membership, with clusters identified using the Louvain algorithm \citep{blondel2008fast} at resolution 1. Only clusters containing more than 4\% of all nodes are shown. In all networks, the size of each node is proportional to its weighted in-degree, representing the number of ratings a Contributor received during that month. The visualisations were created in Gephi 0.10.1.}
    \label{fig:networks}
\end{figure}

Analysis of these rating networks provides valuable insights into the platform's underlying community structure and its evolution over time (see, e.g., \citep{yasseri2023can}). As an illustrative example, Figure~\ref{fig:networks} presents visualisations of the three interaction networks—helpful, somewhat helpful, and unhelpful—constructed from ratings made in January 2023. This figure is intended primarily for illustration, and further analysis is required to fully interpret the structure of the networks. In particular, the unhelpful network (Figure~\ref{fig:networks}(c)) warrants additional attention, since its edges represent negative interactions, yet they were treated as unsigned edges in the current visualisation. 

\section*{Discussion}

Community Notes has become one of the most visible examples of community-based content moderation on a major social media platform. Similar approaches are now being adopted by other platforms \citep{meta2025, tiktok_footnotes2025}, making it increasingly important to understand how such systems operate in practice. Unlike traditional moderation approaches, Community Notes relies on volunteer Contributors to propose contextual Notes and on a rating process designed to identify Notes that are viewed as helpful across different perspectives. This design makes Community Notes an important case for studying the scale, organisation, and limits of community-based moderation.

This paper provides a descriptive investigation of Community Notes during its first four years. Using a large-scale dataset of Notes, ratings, Contributors, cited sources, languages, and interaction networks, we document several core features of the system: linguistic segmentation, widespread use of external sources, concentration of Contributor activity, uneven attention across Posts and Notes, and the relatively low and delayed rate at which Notes reach Helpful status.

Community Notes operates across multiple countries, which is reflected in the linguistic diversity of the Notes. We detected 103 different languages in the Notes. However, Contributor activity appears to remain largely segmented by language. Most Contributors in our dataset write Notes in only one language. Even among Contributors who write in multiple languages, activity is typically concentrated in a single dominant language. These patterns suggest that Community Notes comprises partially separated linguistic communities, in which Contributors primarily participate within a single-language context rather than frequently moving across language groups.

One central design feature of Community Notes is its emphasis on providing sources, and this emphasis appears to be reflected within the Notes. Most Notes include at least one URL, indicating that the system’s encouragement of evidence-based claims is broadly effective. The cited sources are also predominantly neutral in terms of the domain's political orientation. The three most frequently cited websites are X, Wikipedia, and YouTube. This indicates that Contributors frequently rely on other crowd-sourced platforms.

The fourth most cited domain is \url{web3antivirus.io}, an antivirus website focused on cryptocurrency-related content. Further inspection shows that a single highly active Contributor repeatedly wrote Notes on cryptocurrency-related Posts and linked exclusively to this domain. This Contributor also wrote the highest number of Notes in our dataset. Although we cannot determine from the public data whether this activity was automated, the pattern is consistent with potentially automated activity. This case highlights the need for further research on coordinated, promotional, or automated activity within Community Notes.

The thematic analysis reveals that Community Notes activity is concentrated around 
a small number of recurring domains. \textit{Politics \& governance} consistently 
dominates across the entire observation period, while \textit{Health \& medicine} 
was particularly frequent in the early years, driven by COVID-19-related 
misinformation. The relative composition of Note-writing activity shifts over time 
in ways that reflect major real-world events, such as the rise of \textit{War \& 
armed conflict} from late 2023 onward, coinciding with the escalation of the 
Israel-Gaza conflict. These patterns suggest that Community Notes responds 
dynamically to the news cycle.

Turning to Contributor activity, we find that the distributions of Notes and ratings are highly skewed. A small number of Contributors are responsible for a large share of activity.  Similar patterns have been documented in other online crowdsourcing and peer-production systems, including Wikipedia \citep{ortega2008inequality, yasseri2013value}, citizen science projects \citep{sauermann2015crowd}, Reddit \citep{glenski2017consumers}, and Urban Dictionary \citep{nguyen2018emo}. This concentration suggests that openness in formal participation rules does not necessarily translate into equal participation in practice.

One example is the use of the abbreviation “NNN” within Notes, which Contributors use when they believe that a Post does not require additional context. “NNN” refers to “Note Not Needed”. A new Contributor may be unfamiliar with this terminology, suggesting that informal norms and insider language can create barriers to participation.

A similar skewed distribution appears at the Post level. Some Posts receive a large number of Notes, and although the number of ratings generally increases with the number of Notes (Figure~\ref{fig:eda_new}), many highly annotated Posts still do not receive a Helpful Note. This pattern points to a potential limitation of the rating process: additional Contributor activity does not necessarily translate into consensus. One possible explanation, which future work could test using the data released here, is that when multiple Notes compete for attention, ratings may become fragmented across competing Notes. 

These dynamics are reflected in the relatively low rate at which Notes and Posts reach Helpful status. Between January 23, 2021, and January 23, 2025, only 13.55\% of Posts with at least one proposed Note ever received a “Helpful” Note. Across the same period, 87.7\% of all Notes remained in the “Needs More Ratings” category, while only 8.3\% ultimately achieved “Helpful” status and appeared beneath their respective Posts. Even when Notes do become visible, they do so after an average delay of 26 hours, which may occur well after the period of peak visibility for many misleading Posts and thereby reduce the practical impact of the intervention \citep{truong2025delayed}.

The delay between Note creation and Helpful status is, therefore, a central caveat of the current Community Notes system. Prior work suggests several factors that may slow consensus-building in Community Notes, including ideological echo chambers \citep{augenstein2025community} and political polarisation around contentious topics \citep{saeed2022crowdsourced}. Our descriptive results point to an additional possibility: attention is unevenly distributed across Posts and Notes. The fat-tailed distributions shown in Figure~\ref{fig:user-ranks} indicate that Contributor activity is concentrated on a small share of content. Consequently, many Notes may fail to receive enough ratings, or enough ratings from sufficiently diverse Contributors, to reach Helpful status.

Many studies have proposed modifications to Community Notes to improve the system \citep{de2025supernotes, mohammadi2025ai, juncosa2026benefit, costabile2025assessing}. To support further research on this system and on community-based content moderation more broadly, A central contribution of this paper is the release of a four-year dataset covering Community Notes from January 23, 2021, to January 23, 2025. The dataset contains Note IDs, detected languages, extracted topics, cited links, and cited domains for English-language Notes. We also provide monthly interaction networks based on three rating types: helpful, somewhat helpful, and unhelpful. Finally, we release code that allows researchers to reproduce, extend, or adapt the dataset. These resources are intended to support further descriptive and explanatory research on Community Notes, crowdsourced moderation, and the organisation of volunteer-based governance systems on digital platforms.

The present findings should be interpreted in light of several limitations. First, the data used in this paper were downloaded from the Community Notes website on April 7, 2025. Although the dataset covers the period from January 23, 2021, to January 23, 2025, it does not capture Notes or Posts that were deleted before the download date. Community Notes removes deleted Notes and Posts from its publicly released data, so our dataset reflects the public archive available on April 7, 2025, rather than the full historical record of activity over the four-year period. Second, although we detect the language of all Notes, topic modelling and network construction are conducted only for English-language Notes. This limits the extent to which the topic and network analyses can speak to non-English parts of the system. However, the code used to produce these measures is publicly available and can be adapted for use in other languages in future research.

\section*{Methods}

\subsection*{Literature Search and Categorisation}

We conducted a targeted literature search for research on Community Notes. We searched for papers containing the terms ``Birdwatch'' or ``Community Notes'' in their title, abstract, or keywords. We included peer-reviewed articles, preprints, and working papers that were available by November 13, 2025. The resulting papers were categorised by substantive focus and summarised in the literature review section.

\subsection*{Dataset Construction}

Community Notes publicly releases data on Notes and ratings in TSV format. The Notes file contains Note-level metadata, including the Note ID, author ID, referenced Post ID, creation timestamp, the Contributor's classification of the Post as misleading or not misleading, whether trustworthy sources were cited, and the full Note text \citep{communityNotesData}. 

Ratings are released across multiple TSV files, with each row representing a single rating of a Note. These records include the Note ID, the rater's participant ID, the Note's helpfulness, and the reasons selected for the evaluation. The \texttt{helpfulnessLevel} field, introduced on June 30, 2021, replaced the earlier binary \texttt{helpful}/\texttt{notHelpful} label \citep{communityNotesData}.

We downloaded the Notes and ratings datasets on April 7, 2025. From these files, we constructed a curated dataset covering Notes created between January 23, 2021, and January 23, 2025. Because Community Notes removes deleted Posts, along with their associated Notes and ratings, from its publicly released data, any Posts or Notes deleted before April 7, 2025, are not present in the downloaded data. The curated dataset, therefore, reflects the public archive available on the download date rather than the complete historical record of all Notes and ratings created during the study period.

Although additional Community Notes datasets are publicly available, including Note status history and Contributor enrollment records, our analysis focuses on the Notes and ratings datasets. We retain Note IDs and participant IDs so that researchers can link the data released here to other publicly available Community Notes datasets.

\subsection*{Content}

We used the pre-trained FastText language identification model \citep{joulin2017bag} to detect the language of each Note.

An important feature of Community Notes is the inclusion of URLs that provide supporting evidence for a Note’s content. To detect these URLs, we implemented a script that extracted all links from each Note using regular expression (regex) patterns, as listed in Table~S1. Extracted URLs were cleaned by removing whitespaces and surrounding brackets.

For domain-level analysis, we extracted root domains using a secondary regex (Table~S2), which isolated the core domain and returned a list of extracted domains for each Note.
We then normalised the extracted domains to account for platform rebranding, URL shorteners, and regional variants. This step consolidated cases such as the 2023 rebranding of Twitter to X, the use of URL shorteners, and the use of multi-region domain aliases (see Table~S3).

To characterise the thematic content of Community Notes, we developed a two-stage automated classification pipeline that combines large language model (LLM)- based topic induction with iterative human refinement of the resulting taxonomy. In the first stage, we drew a stratified random sample comprising 10\% of the full dataset of English Notes and processed it in batches using GPT-4.1-mini via the OpenAI API \citep{openai2026}. We framed the task as qualitative thematic analysis in the system prompt, instructing the model to assign exactly one topic label per note, reuse existing labels whenever appropriate, and avoid overly fine-grained distinctions (see Supplementary Information). This first analysis resulted in 118 initial topic labels. As is common with LLM-generated taxonomies, the resulting labels included near-duplicates, minor phrasing variations, and semantically overlapping categories. We therefore conducted a manual consolidation step, merging those 118 topics into a first structured taxonomy of 60 categories. This process relied on semantic similarity and thematic grouping; for instance, labels such as \emph{Medical misinformation}, \emph{Health misinformation}, and \emph{Healthcare misinformation} were merged into a single category, \emph{Health \& medical misinformation}. The resulting taxonomy is provided in full in the supplementary materials (Table S4).
To assess the reliability of this taxonomy, we carried out a human annotation study. Four annotators independently labelled a shared set of Community Notes, with each note receiving annotations from three different annotators. Each annotator labelled 201 notes, using a fixed set of the 60 categories.
We evaluated inter-annotator agreement using Fleiss’ $\kappa$, exact match agreement (the proportion of notes where all three annotators agreed), and majority-vote agreement with the LLM-assigned label. Results for this taxonomy are reported alongside those for the refined taxonomy in Supplementary Information.
For the initial 60-category taxonomy, Fleiss’ $\kappa$  was 0.338, indicating fair agreement \citep{landis1977measurement}. Exact match agreement was 17.9\%, and majority-vote accuracy reached 36.2\%. These results reflect the difficulty of distinguishing among a relatively large number of closely related categories, particularly when individual notes span multiple themes (see Supplementary Information).

In response, we undertook a second round of manual refinement.  We then merged related categories. For example,\emph{Health \& medical misinformation} and \emph{Vaccine \& drug misinformation} were combined into \emph{Health \& medicine}, while \emph{Media credibility} and \emph{Source reliability \& bias} were merged into \emph{Media credibility \& bias}. This produced a refined taxonomy of 42 categories, referred to as the second taxonomy (see Table S4 in Supplementary Information)

With this new taxonomy, Fleiss’ $\kappa$ increased to 0.431, corresponding to moderate agreement. Exact match agreement rose to 28.7\%, and majority-vote accuracy increased to 45.9\% (see Figure~S1). 

Following the analysis of the agreement, we applied the refined 42-category taxonomy to the full dataset. The classification prompt was updated in two key ways. First, the complete set of 42 categories was explicitly provided in the system prompt. Second, we allowed up to two labels per note: a required primary label (\emph{label1}) and an optional secondary label (\emph{label2}), enabling the model to capture cases where a note spans multiple themes. Secondary labels were assigned only when clearly justified; otherwise, \emph{label2} was left null (See Supplementary Information for more details).

\subsection*{Interactions}
To examine Contributor interaction dynamics within Community Notes, we constructed monthly networks based on the ratings. 
For the study period, January 2021 to January 2025, we created 49 monthly rating files, each used to generate a corresponding interaction network.
Each network is represented as a directed graph, where nodes correspond to Community Notes Contributors. A directed edge from a rater to a Note writer indicates that the rater evaluated the writer’s Note. Each edge contains three attributes—helpful, unhelpful, and somewhat helpful—representing the number of each respective rating the rater assigned to that Note writer during the month. These complete monthly graphs, referred to as the ``whole networks'', are available for download.

For further analysis, each monthly network was decomposed into three subgraphs, each isolating a different interaction type: a positive subgraph (edges with helpful ratings), a negative subgraph (edges with unhelpful ratings), and a neutral subgraph (edges with somewhat helpful ratings). In each subgraph, edge weights represent the frequency of the corresponding interaction type within that month.

\subsection*{Released Data Files}

For the Notes data, we release Note IDs together with the detected language of each Note, extracted URLs, associated domains, and detected topics for the four-year study period.

For the ratings data, we release 49 monthly TSV files, with each file corresponding to one month in the study period. Each monthly file is accompanied by four GraphML files. The first GraphML file represents the full monthly rating network. In this directed network, an edge from a source Contributor to a target Contributor indicates that the source Contributor rated at least one Note written by the target Contributor during that month. Each edge contains three attributes: \texttt{helpful}, \texttt{unhelpful}, and \texttt{somewhatHelpful}. These attributes represent the number of times the source Contributor assigned each rating type to Notes written by the target Contributor during the month.

The remaining three GraphML files are rating-specific subgraphs of the full monthly network. Each subgraph isolates one rating type, and edge weights indicate the number of times the source Contributor rated the target Contributor's Notes with the corresponding rating during that month. Across all network files, nodes are labelled using participant IDs.

The code used to generate all datasets described above is also released, allowing other researchers to reproduce, adapt, and extend the dataset construction process.

\section*{RESOURCE AVAILABILITY}

\subsection*{Lead contact}
Requests for further information and resources should be directed to and will be fulfilled by the lead contact, Taha Yasseri (taha.yasseri@tcd.ie).

\subsection*{Materials availability}

This study did not generate new, unique reagents.

\subsection*{Data and code availability}
The data and code used in this study are publicly available at~ \url{https://doi.org/10.5281/zenodo.20591253}.

\section*{ACKNOWLEDGMENTS}

The authors would like to thank all the researchers whose work was reviewed in this paper for their valuable comments and feedback, particularly Nicolas Pröllochs. 

\noindent This publication has emanated from research supported in part by grants from Taighde Éireann – Research Ireland under Grant numbers 18/CRT/6049 and IRCLA/2022/3217. TY acknowledges support from Workday, Inc. For the purpose of Open Access, the author has applied a CC BY public copyright licence to any Author Accepted Manuscript version arising from this submission.

\section*{AUTHOR CONTRIBUTIONS}

Introduction, S.M. and T.Y.;
Literature review, S.M., H.C., M.D., S.G., and N.C;
Data merging, K.S.;
Network extraction, S.M., K.S., and M.D.;
Overall statistics, S.M., and H.C.;
Language statistics, N.C., and H.D.;
Source analysis, N.C., and H.D.;
Topic modelling, A.B., N.C., and H.D.;
Network visualisation, S.M.;
Discussion, S.M., and T.Y.;
Data disposition, S.M., A.B.;
Supervision, coordination, and conception, T.Y. and S.M.;
All authors approved the manuscript.

\section*{DECLARATION OF INTERESTS}

The authors declare no competing interests.

\section*{DECLARATION OF GENERATIVE AI AND AI-ASSISTED TECHNOLOGIES}

During the preparation of this work, the authors used ChatGPT 4.1 to improve the writing style of this article. After using this tool, the authors reviewed and edited the content as needed and take full responsibility for the content of the publication.

\bibliography{main}
\newpage
\renewcommand{\thetable}{S\arabic{table}}
\renewcommand{\thefigure}{S\arabic{figure}}

\setcounter{table}{0}  
\setcounter{figure}{0} 
\appendix
\section*{SUPPLEMENTAL INFORMATION for 
\vspace{.5 cm}
\newline {}
\Large From Birdwatch to Community Notes, from Twitter to X:\\ four years of community-based content moderation\\ \newline
{\small Saeedeh Mohammadi, Narges Chinichian, Hannah Doyal, Anna Bertani, Kristina Skutilova, Hao Cui, Michele d’Errico, Siobhan Grayson, Taha Yasseri}
\vspace{.4 cm} 
}

\section{Additional Tables}

\begin{table}[H]
\centering
\caption{URL Extraction Regex Pattern}
\begin{tabular}{|ll|}
\hline
Component                                                                                                 & Matches                              \\ \hline
\texttt{(?:https?|ftp)://\S+} & Protocol-based URLs (HTTP/HTTPS/FTP) \\
\texttt{www\textbackslash.\S+} & \texttt{www}-prefixed domains \\ 
\texttt{[a-zA-Z0-9.-]+\textbackslash.[a-zA-Z]\{2,\}(/\S*)?} & Domain strings with TLDs and optional paths \\\hline
\end{tabular}
\label{tab:regex}
\end{table}

\begin{table}[H]
\centering
\caption{Domain Extraction Regex Pattern}
\label{tab:extraction}
\begin{tabular}{|ll|}
\hline
\textbf{Pattern}                                                                                                 & \textbf{Behavior}                               \\ \hline
\texttt{(?:https?://)?} & Optional protocol \\ 
\texttt{(?:www\textbackslash.)?} & Optional \texttt{www} prefix \\ 
\texttt{([\^{}/]+)} & Captures text until first \texttt{/} (core domain) \\\hline
\end{tabular}
\label{tab:domain_regex}
\end{table}

\begin{table}[H]
\centering
\caption{Domain Normalisation Rules}
\label{tab:normalisation}
\begin{tabular}{|ll|}
\hline
\textbf{Original}                                                                                                 & \textbf{Normalised}                               \\ \hline
\texttt{twitter.com}, \texttt{x.com}, \texttt{t.co} & \texttt{x.com} \\ 
\texttt{youtu.be} & \texttt{youtube.com} \\ 
\texttt{bbc.co.uk} & \texttt{bbc.com} \\
\texttt{*.wikipedia.org} & \texttt{wikipedia.org} \\\hline
\end{tabular}
\label{tab:normalisation_regex}
\end{table}
\section{Topic Classification}
\subsection*{Stage 1: Inductive Topic Discovery}
 
In the first stage, the model was instructed to assign a single topic label to each note. No predefined categories were provided. 
\begin{singlespace}
\begin{tcolorbox}[
  colback=gray!8,
  colframe=gray!50,
  title=\textbf{System Prompt -- Stage 1},
  fonttitle=\bfseries,
  left=6pt, right=6pt, top=6pt, bottom=6pt
]
\begin{verbatim}
You are an expert in qualitative research and thematic analysis.
Your task is to classify Community Notes into stable reusable topics.
 
RULES:
- assign exactly ONE topic per note
- reuse existing topics whenever possible
- avoid creating too many small categories
- keep topic names consistent
- prefer broad but meaningful categories
 
TOPIC LABEL RULES:
- 2-5 words
- concise
- observable themes
- avoid vague labels
- avoid synonyms
 
Return ONLY valid JSON.
 
FORMAT:
[
  {
    "temp_id": 0,
    "topic": "Health misinformation"
  }
]
\end{verbatim}
\end{tcolorbox}
 \end{singlespace}
\subsection*{Stage 2: Deductive Classification with Fixed Categories}
 
In the second stage, the model was provided with a fixed taxonomy of 42 categories derived from Stage 1. Each note was assigned a mandatory primary label (\texttt{label1}) and an optional secondary label (\texttt{label2}), along with a confidence score.
\begin{singlespace}
\begin{tcolorbox}[
  colback=gray!8,
  colframe=gray!50,
  title=\textbf{System Prompt -- Stage 2},
  fonttitle=\bfseries,
  left=6pt, right=6pt, top=6pt, bottom=6pt
]
\begin{verbatim}
You are an expert qualitative researcher.
Your task is to classify Community Notes.

You MUST assign:
- one primary label (label1)
- optionally one secondary label (label2)
 
Allowed categories:
[list of 42 categories]
 
Rules:
- label1 is mandatory
- label2 is optional
- use label2 only when the note clearly belongs to two categories
- if only one category applies, set label2 to null
- DO NOT invent categories
- labels must exactly match one of the allowed categories
- return a confidence score:
    - high
    - medium
    - low
 
Return ONLY valid JSON.
 
Format:
[
  {
    "temp_id": 0,
    "label1": "Health misinformation",
    "label2": "Missing context",
    "confidence": "high"
  }
]
\end{verbatim}
\end{tcolorbox}
\end{singlespace}
\begin{figure}[h!]
    \centering
    \includegraphics[width=1.2 \textwidth, angle=-90]{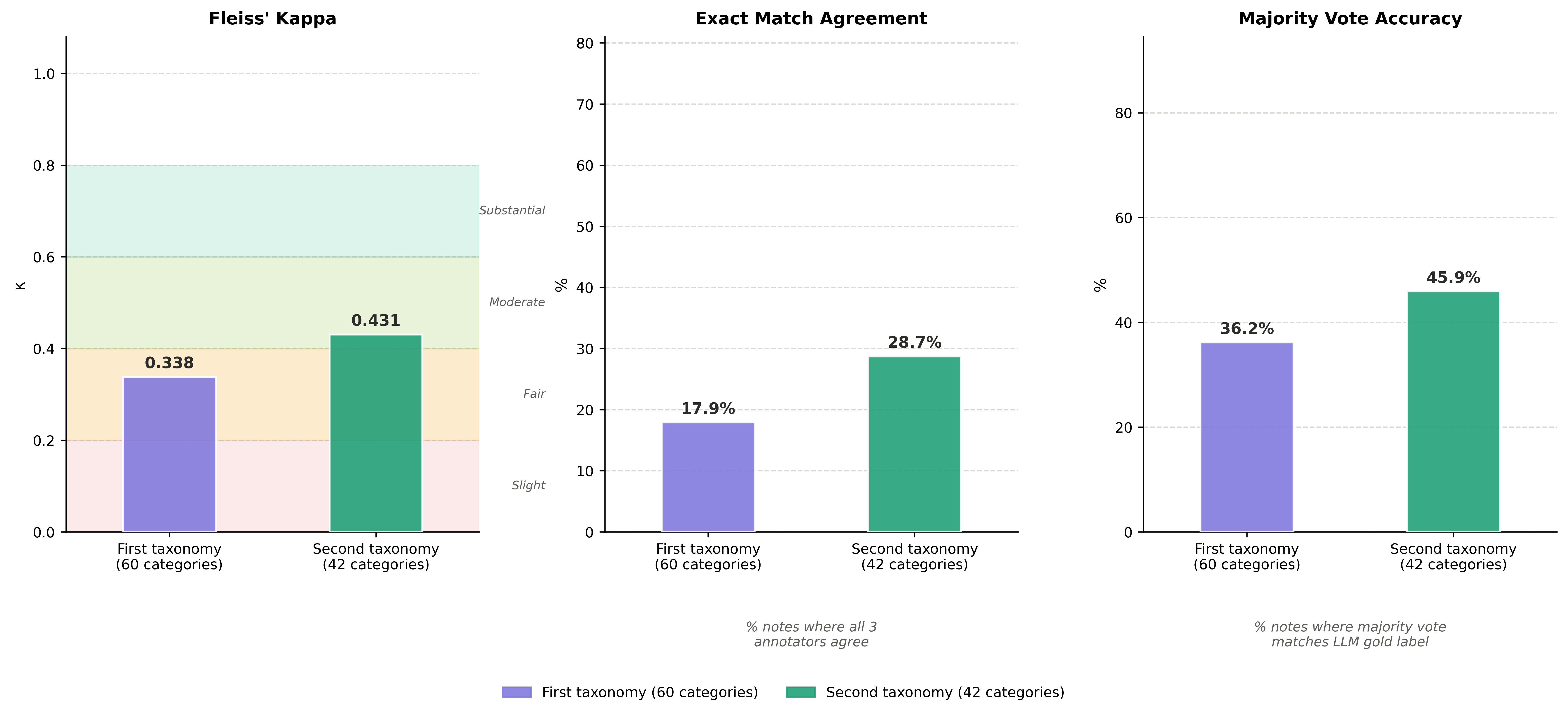}
    \caption{Inter-annotator agreement results across two taxonomy versions. 
    \textbf{Left:} Fleiss' Kappa scores for the first taxonomy (60 categories, $\kappa = 0.338$) 
    and second taxonomy (42 categories, $\kappa = 0.431$), both falling in the \textit{fair-to-moderate} agreement range. 
    \textbf{Centre:} Exact match agreement, defined as the percentage of notes where all three annotators assigned the same label (17.9\% and 28.7\% respectively). 
    \textbf{Right:} Majority vote accuracy, i.e., the percentage of notes where the majority annotation matches the LLM gold label (36.2\% and 45.9\% respectively). 
    Results consistently show improved agreement under the second taxonomy.}
    \label{fig:iaa_results}
\end{figure}

\begin{longtable}{p{5cm}p{5cm}p{5cm}}
\caption{Mapping of original 118 topic to final categories} \\
\toprule
\textbf{Original Label} & \textbf{Intermediate Category} & \textbf{Final Category} \\
\midrule
\endfirsthead
\multicolumn{3}{c}{\tablename\ \thetable{} -- \textit{Continued}} \\
\toprule
\textbf{Original Label} & \textbf{Intermediate Category} & \textbf{Final Category} \\
\midrule
\endhead
\midrule
\multicolumn{3}{r}{\textit{Continued on next page}} \\
\endfoot
\bottomrule
\endlastfoot
Account suspension & Account credibility \& verification & Account integrity \\
Account verification & Account credibility \& verification & Account integrity \\
Account credibility & Account credibility \& verification & Account integrity \\
Account impersonation & Account impersonation \& fraud & Account integrity \\
Account hijacking & Account impersonation \& fraud & Account integrity \\
Adult content misinformation & Adult content misinformation & Adult content \\
Advertising misinformation & Advertising \& marketing misinformation & Advertising \& marketing \\
Product misinformation & Advertising \& marketing misinformation & Advertising \& marketing \\
Art misinformation & Art misinformation & Art \\
Psychological misinformation & Biology \& body science & Health \& medicine \\
Disability misinformation & Biology \& body science & Health \& medicine \\
Biological misinformation & Biology \& body science & Health \& medicine \\
Celebrity misinformation & Celebrity misinformation & Entertainment \& celebrity figures \\
Clickbait and sensationalism & Clickbait and sensationalism & Clickbait \& sensationalism \\
Industrial misinformation & Climate \& environment & Environment \& climate \\
Energy misinformation & Climate \& environment & Environment \& climate \\
Environmental misinformation & Climate \& environment & Environment \& climate \\
Climate misinformation & Climate \& environment & Environment \& climate \\
Community note misuse & Community notes & Community notes \\
Community note usage & Community notes & Community notes \\
Conspiracy misinformation & Conspiracy misinformation & Conspiracy theories \\
Content accuracy & Content accuracy & Content accuracy \\
Content availability & Content availability & Content accuracy \\
Organizational misinformation & Corporate \& business misinformation & Corporate \& business \\
Corporate misinformation & Corporate \& business misinformation & Corporate \& business \\
Crime misinformation & Crime \& law enforcement & Crime \& legal \\
Law enforcement misinformation & Crime \& law enforcement & Crime \& legal \\
Cryptocurrency misinformation & Cryptocurrency misinformation & Cryptocurrency \& scams \\
Linguistic information & Culture \& identity & Culture \& identity \\
Cultural misinformation & Culture \& identity & Culture \& identity \\
Demographic information & Demographic information & Demographics \\
Dietary misinformation & Diet, food \& nutrition & Health \& medicine \\
Food safety misinformation & Diet, food \& nutrition & Health \& medicine \\
Beauty and health misinformation & Diet, food \& nutrition & Health \& medicine \\
Dental health misinformation & Diet, food \& nutrition & Health \& medicine \\
Disaster misinformation & Disaster misinformation & Disaster \\
Labor misinformation & Economic misinformation & Economy \\
Economic misinformation & Economic misinformation & Economy \\
Occupational information & Education \& knowledge & Education \& knowledge \\
Educational misinformation & Education \& knowledge & Education \& knowledge \\
Election misinformation & Election misinformation & Election \\
Misinformation about events & Misinformation about events & Misinformation about events \\
Fictional content & Fictional content & Fictional content \\
Entertainment misinformation & Entertainment misinformation & Entertainment \& celebrity figures \\
Financial misinformation & Financial \& investment misinformation & Finance \& investment \\
Currency misinformation & Financial \& investment misinformation & Finance \& investment \\
Gaming misinformation & Gaming misinformation & Gaming \\
Geographic misinformation & Geographic misinformation & Geography \\
Slander misinformation & Harassment \& defamation & Harassment \& defamation \\
Account harassment & Harassment \& defamation & Harassment \& defamation \\
Illegal content misinformation & Harassment \& defamation & Harassment \& defamation \\
Harassment misinformation & Harassment \& defamation & Harassment \& defamation \\
Medical misinformation & Health \& medical misinformation & Health \& medicine \\
Health misinformation & Health \& medical misinformation & Health \& medicine \\
Healthcare misinformation & Health \& medical misinformation & Health \& medicine \\
Anatomical misinformation & Health \& medical misinformation & Health \& medicine \\
Safety misinformation & Health \& medical misinformation & Health \& medicine \\
Biographical misinformation & History \& biography & History \& biography \\
Historical information & History \& biography & History \& biography \\
Human rights misinformation & Human rights \& social justice & Human rights \& social justice \\
Hate speech misinformation & Human rights \& social justice & Human rights \& social justice \\
Migration misinformation & Immigration \& migration & Immigration \& migration \\
Immigration misinformation & Immigration \& migration & Immigration \& migration \\
Infrastructure issues & Infrastructure \& transport & Infrastructure \& transport \\
Transportation misinformation & Infrastructure \& transport & Infrastructure \& transport \\
Legal information & Legal information & Crime \& legal \\
Ticketing misinformation & Live events \& competitions & Live events \\
Event misinformation & Live events \& competitions & Live events \\
Media misinformation & Media credibility & Media credibility \& bias \\
Fake news & Media credibility & Media credibility \& bias \\
Music industry misinformation & Music industry misinformation & Entertainment \& celebrity figures \\
Weather misinformation & Nature \& wildlife & Environment \& climate \\
Animal misinformation & Nature \& wildlife & Environment \& climate \\
Agricultural information & Nature \& wildlife & Environment \& climate \\
Online scams & Online scams & Cryptocurrency \& scams \\
Personal experience misinformation & Opinion \& satire flags & Satire \& humour \\
Opinion statements & Opinion \& satire flags & Satire \& humour \\
Platform manipulation & Platform manipulation & Platform manipulation \\
Policy disputes & Policy \& governance disputes & Politics \& governance \\
Policy misinformation & Policy \& governance disputes & Politics \& governance \\
Government misinformation & Policy \& governance disputes & Politics \& governance \\
Censorship misinformation & Policy \& governance disputes & Politics \& governance \\
Political misinformation & Political misinformation & Politics \& governance \\
Political statements & Political statements & Politics \& governance \\
Security misinformation & Privacy, copyright \& security & Privacy \& security \\
Privacy concerns & Privacy, copyright \& security & Privacy \& security \\
Copyright misinformation & Privacy, copyright \& security & Privacy \& security \\
Religious misinformation & Religious misinformation & Religion \\
Satire and humor & Satire and humor & Satire \& humour \\
Statistical misinformation & Science \& research & Science \& research \\
Academic misinformation & Science \& research & Science \& research \\
Scientific misinformation & Science \& research & Science \& research \\
Urban planning misinformation & Social norms \& behaviour & Social norms \& behaviour \\
Social misinformation & Social norms \& behaviour & Social norms \& behaviour \\
Concept clarification & Source reliability \& bias & Media credibility \& bias \\
Logical fallacies & Source reliability \& bias & Media credibility \& bias \\
Bias and source reliability & Source reliability \& bias & Media credibility \& bias \\
Satellite and space misinformation & Space \& astronomy & Science \& research \\
Space and satellite misinformation & Space \& astronomy & Science \& research \\
Sport misinformation & Sport misinformation & Sport \\
Digitally altered media & Synthetic \& manipulated media & Synthetic \& manipulated media \\
AI-generated media & Synthetic \& manipulated media & Synthetic \& manipulated media \\
Photographic misinformation & Synthetic \& manipulated media & Synthetic \& manipulated media \\
Symbol misinformation & Synthetic \& manipulated media & Synthetic \& manipulated media \\
Technical misinformation & Technology \& software & Technology \& software \\
Software misinformation & Technology \& software & Technology \& software \\
Accessibility misinformation & Technology \& software & Technology \& software \\
Terrorism misinformation & Terrorism \& extremism & Terrorism \& extremism \\
Bioweapon misinformation & Terrorism \& extremism & Terrorism \& extremism \\
Weapon misinformation & Terrorism \& extremism & Terrorism \& extremism \\
Chemical misinformation & Vaccine \& drug misinformation & Health \& medicine \\
Vaccine misinformation & Vaccine \& drug misinformation & Health \& medicine \\
Drug misinformation & Vaccine \& drug misinformation & Health \& medicine \\
Video content misinformation & Video \& audio misinformation & Video \& audio content \\
Audio content misinformation & Video \& audio misinformation & Video \& audio content \\
War and conflict misinformation & War \& armed conflict & War \& armed conflict \\
Military misinformation & War \& armed conflict & War \& armed conflict \\
Conflict misinformation & War \& armed conflict & War \& armed conflict \\
\end{longtable}

\end{document}